\documentstyle[epsf,seceq,twoside]{ptptex}
\notypesetlogo  %comment in if to eliminate PTPTeX logo
\title{Possible Evidence for a Chiral Particle $D_1^\chi$,\\
Axial-Vector in the $D$ Meson System}
\author{%
Daiki {\sc Ito}, Muneyuki {\sc Ishida}$^*$, Shin {\sc Ishida},
Toshihiko {\sc Komada}$^{**}$\\
Hiroshi {\sc Tonooka} and Kenji {\sc Yamada}$^{**}$ 
}
\inst{%
Research Institute of Quantum Science, 
College of Science and Technology\\
Nihon University, Tokyo 101-0062, Japan\\
$^*$Department of Physics, Tokyo Institute of Technology, 
Tokyo 152-8551, Japan\\
$^{**}$Department of Engineering Science, Junior College Funabashi Campus \\
Nihon University, Funabashi 274-8501, Japan
}
\abst{%
In the covariant level-classification scheme of hadrons proposed recently, 
the existence of ``chiral particles", scalar and axial-vector mesons 
as the partners of the ground-state pseudo-scalar and vector mesons, respectively,
were predicted in the $D$ and $B$ meson systems, realizing a linear representation 
of chiral symmetry. 
In this work 
we reanalyze the $D^*\pi$ mass spectra obtained through the $\Upsilon (4S)$ and $Z^0$ decays
to show a possibility for the $D_1^\chi$.
}
\begin{document}
\maketitle

\setcounter{tocdepth}{4}

\section{Introduction}

The level classification of hadrons had been done rather successfully on the basis of the 
non-relativistic quark model(NRQM) for these 4 decades. However, this non-relativistic 
scheme seems now to be confronted with a serious difficulty: Recently existence\cite{sca,HK} 
of the light-mass
iso-scalar scalar meson to be identified with $\sigma$\cite{rfsigma,rf555}, 
the chiral partner of the $\pi$ meson as a 
Nambu-Goldstone boson, has been accepted widely\cite{2000}. Successively possible existence of the $\kappa$(900)
meson\cite{kappa}, iso-spinor scalar meson, has been pointed out. Then we are able to identify naturally the $\sigma$-meson
nonet\cite{rfmw} with the members $\{ \sigma (600),\kappa (900),a_0(980),f_0(980)  \}$.

The mass values $(\le$ 1GeV) of this scalar nonet are in the region of $q\bar q$ ground states in NRQM,
while there are no seats for scalars in this model. This has aroused recently a hot controversy on what the
quark configuration of $\sigma$ nonet is.

On the other hand we have proposed a new level-classification scheme\cite{rfCLC,u12}, which has a relativistically covariant
framework and in conformity with the approximate chiral symmetry. 
In this new classification scheme the ``chiral states", which are out of the framework of
the NRQM, are expected to exist in the lower mass region, and the $\sigma$-nonet
is naturally assigned as the $(q\bar q)$ relativistic $S$-wave chiral state. In the heavy-light quark meson system
it is expected that the approximate chiral symmetry on the light quark is valid, and the new level-classification scheme
anticipates existence of the chiral states\cite{hl}, scalar mesons and axial-vector mesons, 
in addition to the conventional 
pseudo-scalar and vector mesons, in the ground states. The purpose of this work is to investigate the possibility of
existence of the chiral axial-vector meson $D_1^\chi$ in the $D$ meson system by re-analyzing some experimental data
reported previously.\footnote{
Preliminary results of this work were reported at the conference Hadron 2001 at
Protvino.\cite{yamada}
}

\section{Experimental Data and Project for Reanalysis}

\hspace*{-0.8cm} ({\it Experimental data to be analyzed})

As the experimental data\cite{CLEO,DELPHI} to be re-analyzed we choose the mass spectra of $D^{*+}\pi^-$ system
obtained through the two processes,
\begin{eqnarray}
{\rm CLEO\ II} &:& e^++e^-\rightarrow \Upsilon (4S) \rightarrow D^{*+}\pi^- + \cdots ,\ \ \label{eq2a}\\
{\rm DELPHI} &:& e^++e^-\rightarrow Z^0 \rightarrow D^{*+}\pi^- + \cdots ,\ \ \label{eq2b}
\end{eqnarray}
which have at least a necessary statistical accuracy deserved to meaningful analysis.
We shall take the two kinds of CLEO data on the process Eq.~(\ref{eq2a}),
CLEO (a) and CLEO (b); the former are the restricted ones so as to stress relatively 
the conventional axial-vector meson $D_1^0$, while the latter are all the relevant data.

As the irrelevant backgrounds, which come from the other than decay processes of 
the relevant resonant particles, the CLEO group set up the formula
\begin{eqnarray}
B.~G.~ &=& \alpha (\Delta M)^\beta \times 
{\rm exp}(-\gamma_1(\Delta M)-\gamma_2(\Delta M)^2-\gamma_3(\Delta M)^3)\ ,
\ \  \label{eq3}\\
\Delta M &\equiv& M(D^*\pi)-(m_{D^*}+m_\pi ),\ \ \nonumber
\end{eqnarray}
with fitting parameters $\alpha ,\ \beta ,\ \gamma_1,\ \gamma_2 ,\ \gamma_3$ , while the DELPHI group
applies the similar formula to Eq.~(\ref{eq3}) with fitting parameters 
$\alpha ,\ \beta ,\ \gamma_1$, setting $\gamma_2 =\gamma_3 =0$.
We shall apply the same formula to Eq.~(\ref{eq3}) with parameters 
$\alpha ,\ \beta ,\ \gamma_1,\ \gamma_2,\ \gamma_3$ for the backgrounds,
taking into account the possible effects of intermediate $D_1^\chi$ production.

\hspace*{-0.8cm} ({\it Relevant intermediate resonances})

The conventional three $(c\bar q)-P$ wave mesons may contribute, as intermediate states, 
to the final $D^{*+}\pi^-$ system with the respective angular momenta $l=0,1,2$ as
\begin{eqnarray} 
D_1^* &\rightarrow& D^{*+} + \pi^-\ \ (S-{\rm wave})\ ,   \label{eq4a} \\
D_1 &\rightarrow& D^{*+} + \pi^-\ \ (D-{\rm wave})\ ,   \label{eq4b} \\
D_2^* &\rightarrow& D^{*+} + \pi^-\ \ (D-{\rm wave})\ ,   \label{eq4c} 
\end{eqnarray}
where $D_1^*,\ D_1$ and $D_2^*$ have, respectively, ${}^{j_q}L_J={}^{1/2}P_1,\ {}^{3/2}P_1$ and ${}^{3/2}P_2$
$( {\mib j}_q={\mib S}_q+{\mib L}$ is the total amgular momentum of the light quark),
and the respective partial wave states in Eqs.~(\ref{eq4a},\ref{eq4b},\ref{eq4c}) are deduced from the heavy quark symmetry (HQS).
In this work, a possible contribution from, in addition to the above conventional resonances, the chiral axial-vector
meson $D_1^\chi$ is taken into account as 
\begin{eqnarray}
D_1^\chi & \rightarrow & D^{*+} + \pi^- \ (S-{\rm wave})\ , \ \ 
\end{eqnarray}
where we inferred that the $S$-wave decay is dominant because of small $Q$-value.

Out of the three resonances in Eqs~(\ref{eq4a},\ref{eq4b},\ref{eq4c}) the $D_1$ and the $D_2^*$,
decaying in the $D$-wave,
have comparatively a small decay width $\sim 20$ MeV, while $D_1^*$,
decaying in the $S$-wave, has a large width $\sim 300$ MeV,
reflecting the physical situation that all three resonances have small $Q$-value and that the width contains a
kinematical factor $\Gamma \propto P^{2l+1}$ ($P$ being relative momentum of the final system).

Because of its large width the contribution of $D_1^*$ is difficult to be discriminated from the backgrounds and it
was treated as being included into the background in the original analysis by the experimental groups.
In this work we shall follow this prescription.

Thus, in order to analyze the relevant mass spectra, we shall take into account the three resonances,
$D_1,\ D_2^*$ and $D_1^\chi$.   

\hspace*{-0.8cm} ({\it Method of analysis})

We shall apply the VMW method in our relevant case, where the absolute amplitude squared is given by
\begin{eqnarray}
|M(s)|^2 &=& \sum_i  | r_i \Delta_i (s) |^2 + {\rm B. G. }\ , \label{eq8}\\
    \Delta_i & \equiv & \frac{-m_i\Gamma_i}{s-m_i^2 + i m_i\Gamma_i}\ .\nonumber
\end{eqnarray}
Here the summation on the suffix $i$ goes over the resonant particles $D_1,\ D_2^*$ and $D_1^\chi$,
$m_i(\Gamma_i)$ are their masses (widths), and $r_i$ represent their production strength.

The relevant mass spectrum is given by
\begin{eqnarray}
\Gamma (s) &=& \frac{1}{2\sqrt s} \int 
\frac{d^3{\mib P}_{D^*} d^3{\mib P}_\pi }{(2\pi )^3 2E_{D^*}(2\pi )^3 2E_\pi}\nonumber\\
  & \times & (2\pi )^4\delta^{(4)} (P-P_{D^*}-p_\pi ) | M(s) |^2 \ ,\label{eq9}
\end{eqnarray}
where $s\equiv -(P_\mu )^2$, and $P_\mu$ being the total 4-momentum of relevant system.

\hspace*{-0.8cm} ({\it Project for applying $\chi^2$-analysis})

In order to obtain the least-$\chi^2$ solution on our relevant parameters, we set up the restrictions
on the values of resonance masses and widths as is collected in Table I.

\begin{table}
\caption{Restricted regions of values of masses and widths (MeV).} 
\begin{center}
\begin{tabular}{c|ccc}
\hline
       & $D_1^\chi$ & $D_1$ & $D_2^*$ \\
\hline
$m$ & 2295-2320 & 2400-2450 & 2420-2480 \\
$\Gamma$ & 10-60 & 10-35 & 10-100 \\
\hline
\end{tabular}
\end{center}
\end{table}

The regions for $D_1$ and $D_2^*$ are determined properly from the center values 
and errors of respective resonances reported in PDG table. Concerning a possible
candidate for the chiral particle $D_1^\chi$, some excess structure over backgrounds 
in the regions centered at $m\sim 2300$ MeV, (which is seen commonly through all
the three experimental data of CLEO (a) and (b), and of DELPHI) is identified as being
due to production of $D_1^\chi$. The restrictions on $D_1^\chi$ come simply from 
inspection of this structure.

We shall perform the two kinds of $\chi^2$-fitting of the mass spectra, the one with $D_1^\chi$
and the other without $D_1^\chi$ and compare with their results.

\section{Results of Analysis} 

The results of our analysis in both the cases, with $D_1^\chi$ and without $D_1^\chi$,
are shown in Fig.~1:\ \ \ 
The fitted curves in comparison with the data of CLEO (b), DELPHI and CLEO (a) are given,
respectively, in Fig.~1(a), (b) and (c) $[$ (d), (e) and (f)  $]$
in the case with $D_1^\chi\ [$ without $D_1^\chi ]$. 
The respective contributions from the $D_1^\chi$,
$D_1,\ D_2^*$ and B. G. are also shown.   

Actually we have made our analysis in the two steps(, taking into account the difference of statistical accuracy
of respective data): First fitting CLEO (b) and DELPHI commonly and secondly fitting CLEO (a) with the 
fixed values of resonance mass and width obtained through the first fitting.

\begin{figure}
\epsfysize=13. cm
  \centerline{\epsffile{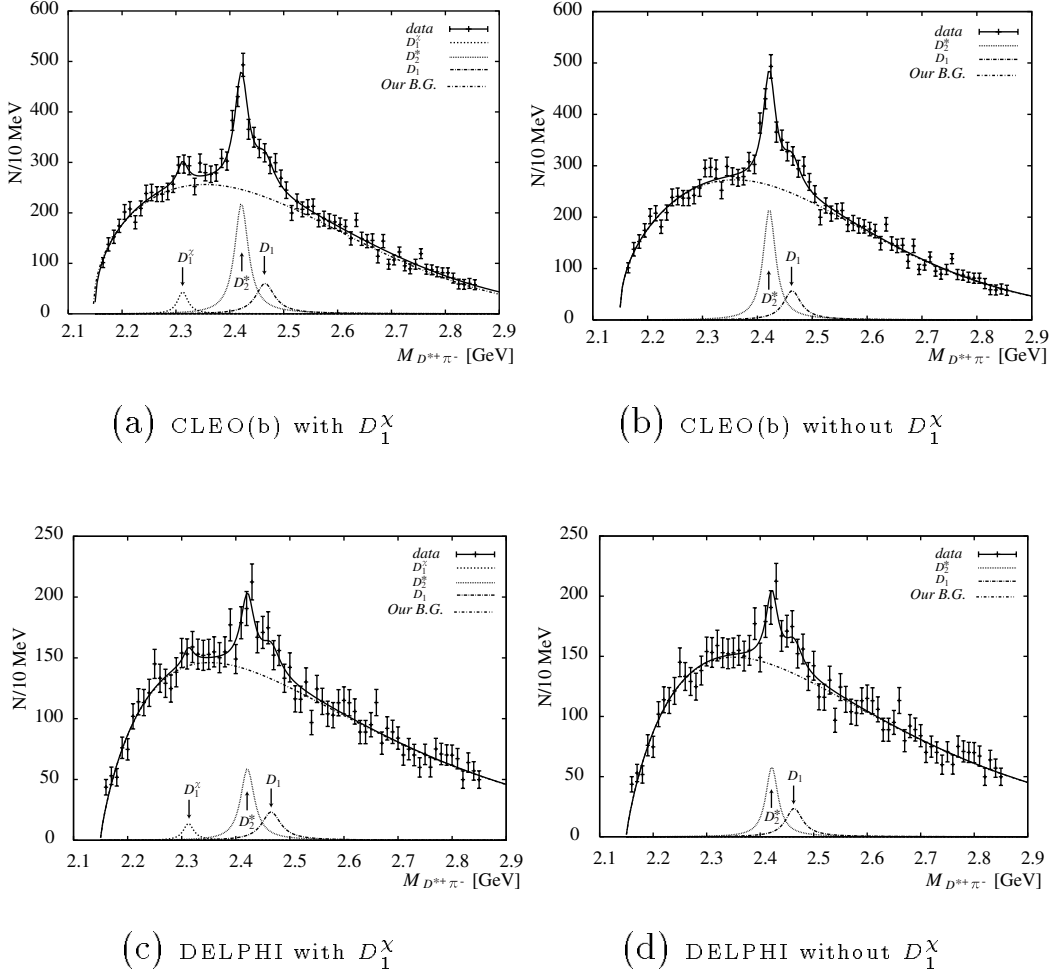}}
\caption{Fitted curves in the case with (without) $D_1^\chi$ and experimental data
of CLEO (b) and DELPHI are shown, respectively, in (a) and (c) ( (b) and (d) ). }
\end{figure}

\begin{figure}[t]
\epsfysize=6.5 cm
  \centerline{\epsffile{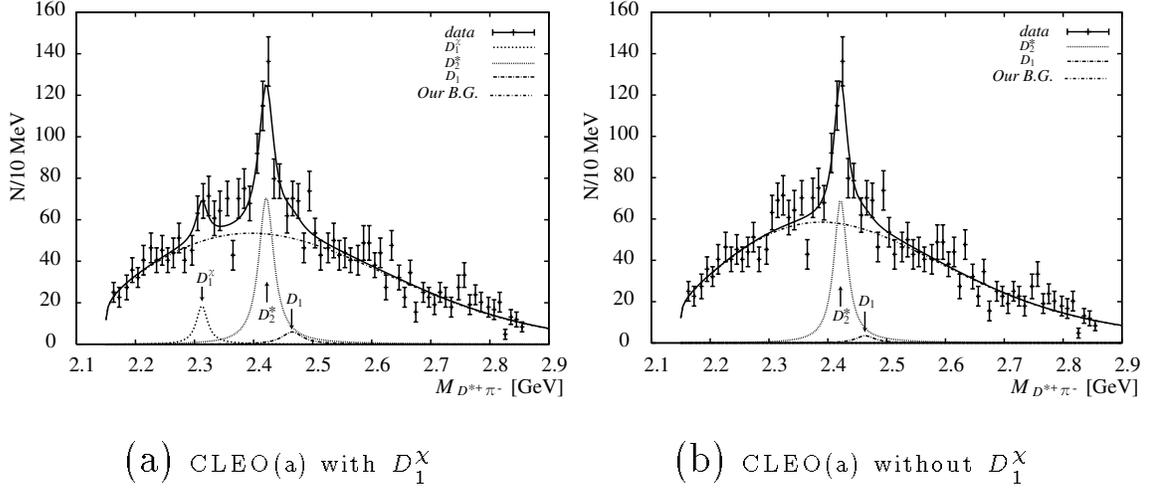}}
\caption{Fitted curves in the case with (without) $D_1^\chi$ are, for reference, given in comparison
with experimental data, CLEO(a), where the values of mass and width of relevant resonances are determined
in the analysis of CLEO(b) and DELPHI. }
\end{figure}

The obtained values of mass and width of the relevant resonances are collected
in Table II.
\begin{table}
\caption{Fitted values of mass and width of resonances (in MeV)}
\begin{center}
\begin{tabular}{c|ccc}
\hline
     &  $D_1^\chi$  &  $D_1$   &   $D_2^*$    \\  
\hline
$m$ & 2312 & 2421 & 2465 \\
$\Gamma$ & 23.03 & 30.73 & 42.54 \\
\hline
\end{tabular}
\end{center}
\end{table}
The values of $D_1$ and $D_2^*$ are close to those given in PDG table.

In Table III the values of $\chi^2$ and of 
$\tilde\chi^2(\equiv \chi^2/{\rm No.\ of\ data\ points-No.\ of\ param.})$ in the 
respective steps of fitting are also given.

\begin{table}
\caption{Values of $\chi^2$ and $\tilde\chi^2$}
\begin{tabular}{l|c|c|c|c|c|c}
\hline
     &  \multicolumn{3}{c|}{(A) with $D_1^\chi$}  &   \multicolumn{3}{c}{(B) without $D_1^\chi$}  \\  
\hline
     & CLEO (b) & DELPHI & CLEO (a) & CLEO (b) & DELPHI & CLEO (a) \\
\hline
 $\tilde\chi^2$ 
     & \multicolumn{2}{c|}{$\frac{\displaystyle 110}{\displaystyle (140-20)}=0.923$} 
     & $\frac{\displaystyle 80.6}{\displaystyle (70-8)}=1.30$ 
                        & \multicolumn{2}{c|}{$\frac{\displaystyle 119}{\displaystyle (140-16)}=0.957$}
          & $\frac{\displaystyle 85.5}{\displaystyle (70-7)}=1.36$ \\ 
\hline
$\chi^2$  & 59.51 & 51.28 &     &   66.41  &  52.25   &     \\
\hline
\end{tabular}
\end{table}

Through the inspection of fitted curves in Fig.~1,
we may conclude that our results give some evidence for existence of $D_1^\chi$, 
although there is no significant improvement on the reduced $\chi^2$ given in Table III.

\section{Concluding Remarks}

We have made reanalysis of the $(D^{*+}\pi^-)$ mass spectra obtained through the processes Eqs.~(\ref{eq2a})
and (\ref{eq2b}), by applying the formulas Eqs.~(\ref{eq8}) and (\ref{eq9}). 
As the intermediate resonant particles we have taken into account the $D_1^\chi$,
a new chiral axial-vector meson outside of the conventional level classification scheme.
As a result we have obtained a possible evidence for $D_1^\chi$ with $m=2312$ MeV and $\Gamma = 23.0$ MeV.
The obtained values of reduced $\chi$ square,
$\tilde\chi^2\equiv \chi^2/({\rm No.\ of\ data\ points- No.\ of\ param.})$, are\\
$\tilde\chi^2=110/(140-20)=0.923$ for the case with $D_1^\chi$
and 
$119/(140-16)=0.957$ for the case without $D_1^\chi$, respectively.
However, the statistical accuracy of the relevant data is very poor.
It is required to have the more accurate data in order to get a definite conclusion. 

A possible evidence for another chiral particle $B_0^\chi$, scalar meson in the $B$ system, is also shown
by reanalysis of the $B\pi$ mass spectra through the decay process of $Z^0$-boson. 
It will be given in a separate paper.\cite{B0chi}

\acknowledgement

The authors should like to express their gratitude to Professors, K. Takamatsu and T. Tsuru,
for their useful discussions and encouragement.

\end{document}